\documentclass[%
 aip,
 amsmath,amssymb,
 reprint,%
]{revtex4-1}

\usepackage{dcolumn}
\usepackage{bm}

\usepackage[utf8]{inputenc}
\usepackage[T1]{fontenc}
\usepackage{mathptmx}
\usepackage{etoolbox}

\usepackage{algcompatible}
\usepackage{amsmath}
\usepackage{amsthm}
\usepackage{bm}
\usepackage{braket}
\usepackage{subfigure}
\usepackage[dvipdfmx]{graphicx}
\usepackage{float}
\usepackage{url}
\usepackage{xcolor}
\usepackage{multirow}
\usepackage[colorlinks=true,urlcolor=blue,citecolor=blue,linkcolor=blue]{hyperref}
\usepackage{cleveref}
\usepackage{here}
\usepackage{algorithm}
\usepackage{physics}
\usepackage[noend]{algpseudocode}
\usepackage{comment}

\crefname{figure}{Fig.}{Figs.}
\crefname{equation}{Eq.}{Eqs.}
\crefname{section}{Sec.}{Sec.}
\Crefname{figure}{Figure}{Figures}
\Crefname{equation}{Equation}{Equations}
\Crefname{section}{Section}{Sections}

\makeatletter
\def\@email#1#2{%
 \endgroup
 \patchcmd{\titleblock@produce}
  {\frontmatter@RRAPformat}
  {\frontmatter@RRAPformat{\produce@RRAP{*#1\href{mailto:#2}{#2}}}\frontmatter@RRAPformat}
  {}{}
}%
\makeatother
\begin{document}

\preprint{AIP/123-QED}

\title[Mechanically-intermixed indium superconducting connections for microwave quantum interconnects]{Mechanically-intermixed indium superconducting connections \\ for microwave quantum interconnects}

\author{Yves Martin}
\author{Neereja Sundaresan}
\author{Jae-woong Nah} 
\author{Rachel Steiner} 
\author{Marco Turchetti} 
\author{Kevin Stawiasz} 
\author{Chi Xiong}
\author{Jason S. Orcutt}
\email{jsorcutt@us.ibm.com}
\affiliation{IBM Quantum, IBM T.J. Watson Research Center, Yorktown Heights, New York 10598, USA}

\date{\today}

\begin{abstract}
Superconducting coaxial cables represent critical communication channels for interconnecting superconducting quantum processors. Here, we report mechanically-intermixed indium joins to aluminum coaxial cables for low loss quantum interconnects. We describe an ABCD matrix formalism to characterize the total resonator internal quality factor ($Q_i$) and any contact ($R_{cont}$) or shunt resistance ($R_{shunt}$) associated with the mechanically-intermixed indium joins. We present four resonator test systems incorporating three indium join methods over the typical frequency range of interest (3-5.5GHz) at temperatures below $20mK$. We measure high internal quality factor aluminum cables ($Q_i = 1.55 \pm 0.37 x 10^6$) through a push-to-connect indium join of the outer conductor that capacitively couples the inner conductor for reflection measurements. We then characterize the total internal quality factors of modes of a cable resonator with a push-to-connect superconducting cable-splice at the midpoint to find mean $Q_i = 1.40 x 10^6$ and $Q_i = 9.39 x 10^5$ for even and odd-modes respectively and use an ABCD matrix model of the system to extract $R_{cont} = 6x10^{-4} \Omega$ for the indium join of the inner conductor. Finally, we demonstrate indium press-mold cable-to-chip connections where the cable-to-chip join is placed at a current node and voltage node through varying on-chip waveguide lengths with mean $Q_i = 1.24 x 10^6$ and $Q_i = 1.07 x 10^6$ respectively to extract $R_{cont} = 8.5x10^{-4} \Omega$ and $R_{shunt} = 1.3x10^7 \Omega$ for the interface. With these techniques, we demonstrate a set of low-loss methods to join superconducting cables for future quantum interconnects.
\end{abstract}

\maketitle

As superconducting quantum processing units (QPUs) have reached the scale to hundreds of physical qubits \cite{ibm_quantum_systems}, including the recent demonstration of a 1121 qubit processor \cite{gambetta2023ibm}, new technologies are needed to connect multiple QPUs together to build a practical, error-corrected computational system with tens of thousands to millions of physical qubits \cite{bravyi2022future,awschalom2021development,awschalom2022roadmap}. Following the terminology of Bravyi et.al\cite{bravyi2022future}, the required chip-to-chip quantum interconnects can be roughly categorized into three types: m-couplers that connect proximate chips within a single processor package \cite{gold2021entanglement}, l-couplers that connect separately packaged QPUs within a single cryogenic environment \cite{kurpiers2018deterministic,zhong2021deterministic,burkhart2021error,yan2022entanglement,niu2023low,qiu2023deterministic}, and t-couplers that connect QPUs across isolated cryogenic environments \cite{krastanov2021optically,ang2022architectures}. Of these coupler types, the l-couplers are particularly relevant for near term technology impact by extending the computational system beyond the bottlenecks of control signal routing for a single QPU payload, which are still present for m-coupled systems, while avoiding the challenging physics problems associated with interfacing superconducting qubits to optical photons, which must be solved to enable t-coupled systems.

A key physical challenge to implement l-coupler quantum interconnects is the realization of a low loss microwave channel that connects on-chip transmission lines of planar superconducting circuits via superconducting coaxial cables. This requires both a low-loss cable and a low-loss method of attaching the cable to a chip. Initial reports of superconducting cable quality factors of order $10^5$ for LD-PTFE dielectric-based NbTi cables \cite{kurpiers2017characterizing,burkhart2021error,zhong2021deterministic}. Fundamentally, this channel quality factor is sufficient to realize state transfer infidelities below $1\%$ \cite{malekakhlagh2024enhanced}, but recent literature reports even higher internal quality factors above $10^6$ in Al cables \cite{niu2023low,qiu2023deterministic}. These reports are generally consistent with an earlier study of the PTFE microwave (18.92 GHz) loss tangent of $2.3 x 10^{-6}$  at cryogenic conditions (28 K)\cite{jacob2002microwave} considering the variability in material purity and the lower effective electric field participation in the low density material used for these cables. Previous reports of cable-to-chip attach for quantum interconnects have relied upon wirebond connections between the chip and both inner and outer conductors of the cables \cite{zhong2021deterministic,niu2023low,qiu2023deterministic}. Characterization of this connection estimated a contact resistance to the inner conductor of $R_{cont} = 0.38\Omega$ \cite{zhong2021deterministic} for total link internal quality factors of approximately $10^4$. Refinements of this technique to place the wirebond connection at a current node for a narrow frequency range of resonant modes by tuning the on-chip transmission line length have achieved $8.1 x 10^5$ internal quality factor \cite{niu2023low}.

In this work we demonstrate the use of mechanically intermixed indium (cold-welding) as a low-loss, robust, and scalable method of interfacing superconducting coaxial lines with planar superconducting circuits to create cable resonator systems with internal quality factors ($Q_i$) above $10^6$ across a broad frequency range relevant to quantum computing of 3-5.5GHz. We demonstrate a push-to-connect indium join for the outer conductor of coaxial cables for accurate characterization of cable resonators in a reflection geometry. We then report a "splice" of inner and outer conductors using push-to-connect indium joins as an alternative to the wirebond connnections reported in \cite{qiu2023deterministic} or proprietary mechanical joints as reported \cite{renger2023cryogenic}. Finally, we characterize a press-molded  attach of the superconducting cable to an on-chip superconducting co-planar waveguide. For all cases, we model the experimental data with microwave-circuit equivalents to estimate component performance parameters. Together, these techniques eliminate the need for delicate, bench-top wirebonding to create high-quality quantum channels and bring us one step closer to realizing large QPUs connected by meter length-scale quantum links in a single cryogenic environment. 

The superconducting coaxial line used in all experiments for this paper has an aluminum inner and outer conductor, a low-loss LD-PTFE dielectic, and $\sim$2.05mm outer diameter - supplied by Nanjing HMC System Co.,Ltd. This cable has also been used previously in prior reported quantum interconnect experiments reporting internal quality factors of order $10^6$ \cite{niu2023low,qiu2023deterministic}. The methods presented in this paper extend beyond this specific cable and have been tested internally with cables comprising a variety of superconducting metals from a variety of vendors, but for clarity we present data only from this example superconducting line. 

\begin{figure*}
\centering
\includegraphics[width=1.0\textwidth]{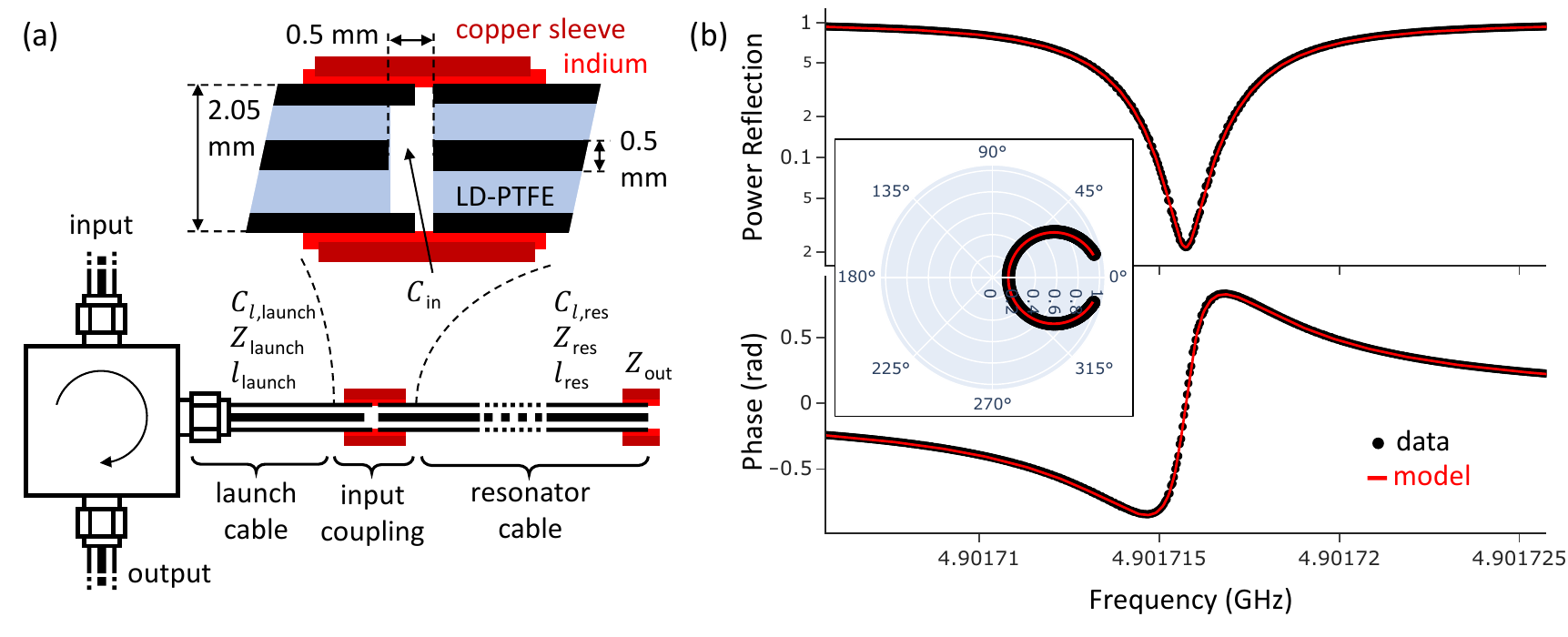}
\caption{(a) Experimental setup showing measurement of $S_{11}$ of the network calculated in Equation \ref{eqn:cable} using a circulator. The launch cable consists of a short coaxial line (copper or NbTi) with a SMA connector on one end. At the other end, the center-pin is recessed by $\sim$ 0.5 mm (establishing $C_{in}$). A copper tube with thin indium coating holds together the launch cable and the resonator-under-test securely, and provides ground connection between the two. In this depiction, the resonator-under-test is the bare cable. See main text for details on cable preparation. Parameters $C_{l}$, $Z$, and $l$ for launch and resonator under test sections, $C_{in}$, and $Z_{out}$ are used for ABCD model fitting. (b) Exemplary fit of power reflection and phase as a function of frequency (inset: complex field polar plot) with Eqn. \ref{eqn:cable} for a cable measured in the configuration of (a) from Fig. ~\ref{fig:Qint_bare}. Fit parameters for the $n=41$ mode at 4.902 GHz were $l=0.9999 m$, $C_{in}=3.602 fF$, $Q_i = 1.556 x 10^6$, a phase offset, and reflection amplitude normalization. This corresponds to an external quality factor of $2.093 x 10^6$ and a total quality factor of $8.926 x 10^5$.}
\label{fig:reflection_measurement_diagram}
\end{figure*}

To measure the internal quality factor of the cables and assemblies, we adopt the reflection measurement setup shown in Figure \ref{fig:reflection_measurement_diagram}. While there are many potential microwave configurations possible for such measurements \cite{mcrae2020materials}, the reflection measurement of a single port through a circulator enables a single physical port at a well defined field point of the resonator. For the experiments presented here, the circulator and cable assembly are anchored to the mixing chamber plate of a dilution refrigerator. RF signals input to the circulator are attenuated by 70 dB of fixed attenuators across the various thermal stages and 10.4 dB of cable loss with no additional filtering. The output of the circulator is fed into an isolator,  which connects to a standard HEMT amplifier at 4K via a NbTi cable. Although prior work has analyzed similar measurements \cite{castellanos2007widely,schuster2007circuit,rieger2023fano}, here we adopt a modeling framework based on ABCD matrices \cite{pozar2012microwave,chen2022scattering} that enables simple fitting of internal resonator parameters and prediction of system performance across multiple resonant modes with a single physical model.

In these models, the 2x2 matrix with defined parameters $[[A,B],[C,D]]$ relates the output voltage and current ($[[V_2],[I_2]]$) to the input voltage and current ($[[V_1],[I_1]]$). This relation can then be converted to the experimentally measured reflection parameter $S_{11}$ for the resonator from a system with nominal impedance $Z_0$ by the following equation \cite{pozar2012microwave}:

\begin{equation}
S_{11} = \frac{A + B/Z_0 - C Z_0 - D}{A+B/Z_0 + C Z_0 + D}
\end{equation}

In the ABCD matrix formalism, the 2x2 matrices for each component in a microwave network can be multiplied to find the overall response to the input or output port for frequency $f$. Using the definitions in Figure \ref{fig:reflection_measurement_diagram}, we can define the simple case of a launch transmission line, whose inner conductor is capacitively coupled to the inner conductor of the cable resonator under test as follows:

\begin{equation}
\begin{bmatrix} A & B \\ C & D \end{bmatrix}
=
\bm{\mathit{T_{launch}}}
\begin{bmatrix} 1 & \frac{-j}{2\pi f C_{in}} \\ 0 & 1 \end{bmatrix}
\bm{\mathit{T_{res}}}
\begin{bmatrix} 1 & Z_{out} \\ 0 & 1 \end{bmatrix}
\label{eqn:cable}
\end{equation}

Each transmission line segment ABCD matrix $\bm{\mathit{T_x}}$ is calculated from its impedance $Z_x$, length $l_x$, and complex propagation constant $\beta_x = j \alpha_x / 2 + 2 \pi f \sqrt{\epsilon_{R,x}} / c$, where $\alpha_x$ is the transmission line power attenuation coefficient, $\epsilon_{R,x}$ is the transmission line relative permittivity, and  $c$ is the vacuum speed of light:

\begin{equation}
\bm{\mathit{T_x}} = \begin{bmatrix} \cos(\beta_{x} l_{x}) & j \sin(\beta_{x} l_{x}) \\ \frac{j}{Z_{x}}\sin(\beta_{x} l_{x}) & \cos(\beta_{x} l_{x}) \end{bmatrix}
\end{equation}

The transmission line relative permittivity $\epsilon_R$ and attenuation coefficient $\alpha$ can be related to more commonly used parameters of internal quality factor $Q_i$, capacitance per unit length $C_{l}$, and impedance $Z_0$:

\begin{equation}
    \epsilon_R = (c C_{l} Z_0)^2
\end{equation}

\begin{equation}
    \alpha = \frac{2 \pi f \sqrt{\epsilon_R} }{c Q_i}
\end{equation}

The physical input coupling capacitance used for the model presented here can also be related to the more commonly used abstract resonator parameter of the external quality factor, $Q_e$, for any given mode order $n$, to additionally define the total quality factor, $Q_t$:

\begin{equation}
    Q_e = \frac{n}{8\pi(f C_{in} Z_0)^2}
\end{equation}
\begin{equation}
    Q_t = \frac{1}{1 / Q_i + 1 / Q_e}
\end{equation}

\begin{figure}
\includegraphics[width=0.48\textwidth]{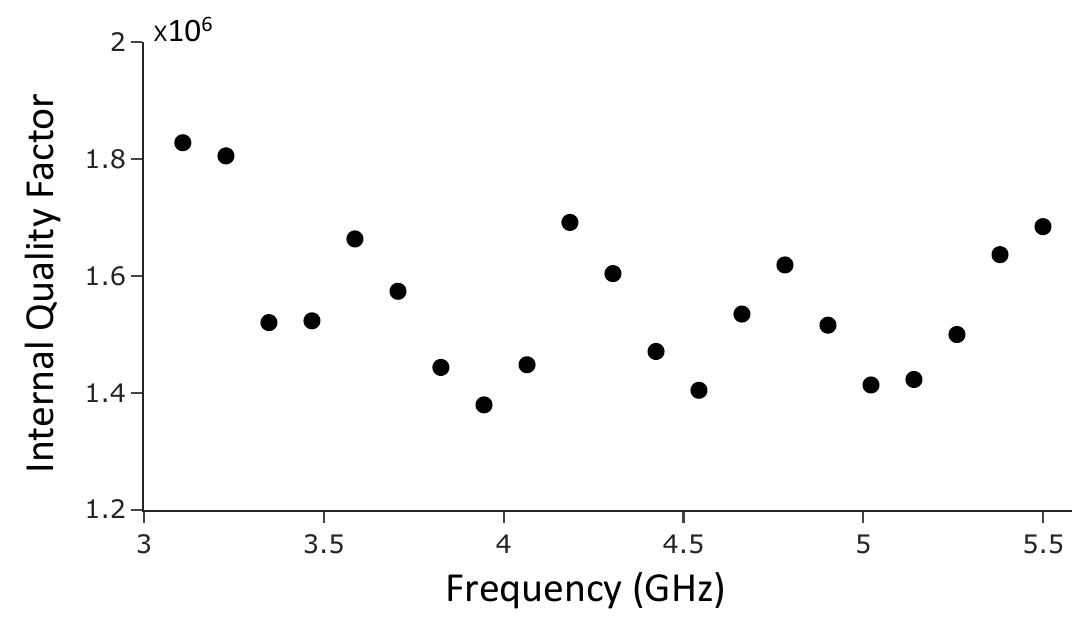}
\caption{Fit internal quality factors for the bare cable resonator of the 1m length aluminum superconducting coaxial cable under study. Mean internal quality factor over the range of 3-5.5 GHz is $1.55 x 10^6$ with a standard deviation of $1.24 x 10^5$}
\label{fig:Qint_bare}
\end{figure}

In this notation, the fundamental mode of the resonator system corresponds to $n=1$. Using this model we show an exemplary fit of a cable resonance in Fig. \ref{fig:reflection_measurement_diagram}(b) and in Fig. \ref{fig:Qint_bare} plot the fitted $Q_i$s over a wider frequency range showing high $Q_i$ throughout. The mean internal quality factor is $Q_i = 1.55 \pm 0.37 x 10^6$ with $3\sigma$ error bars generated from the statistical variation of internal quality factors across longitudinal mode orders. Converted to a cable attenuation coefficient per unit length, this loss rate is equivalent to $\alpha \approx 0.39 dB/km$, or within a factor of three of the lowest-loss single-mode optical fiber \cite{kawaguchi2015ultra}. We can also use this data to place an approximate bound on the loss tangent of PTFE. Given that the equivalent loss tangent mean equivalent of the quality factor is $6.45 x 10^{-7}$ and the effective relative permittivity ($\epsilon_R$) from the model that fits the measured cable free spectral range is 1.55, we can correct the loss tangent relative to a bulk PTFE relative permittivity of 2.05 \cite{krupka2016measurements} to estimate a loss tangent of $8.53x10^{-7}$, which is approximately a factor of four lower than prior estimates at a higher frequency and temperature than these reported results \cite{jacob2002microwave}.

For the push-to-connect indium attach of the outer conductor of the resonator cable under test depicted in Fig. ~\ref{fig:reflection_measurement_diagram}, we first prepare a launch cable (copper or NbTi for rigidity) to have a standard SMA connector on one end to connect with the circulator and a simple notch on the other end to recess the center-conductor by approximately 0.5mm to establish weak coupling ($C_{in} \sim 3fF$) with the aluminum cable. To join the launch cable with the aluminum cable, we prepare the outer ground conductors of both with a thin layer of 99.995\% pure indium, applied using ultrasonic soldering at 160\textdegree C. Then we can press-fit both these cables into a copper tube prepared with a thin coating of indium inside, the mechanical friction from the pressing action is sufficient for strong indium-indium bonding. A similar tube is press-fit on the end of the aluminum cable. These tubes are prepared well in advance of final usage, by first filling a copper tube with indium using a combination of flux and heat, and then using a drill-press to precisely core out the indium to desired thickness. 

\begin{figure}
\includegraphics[width=0.48\textwidth]{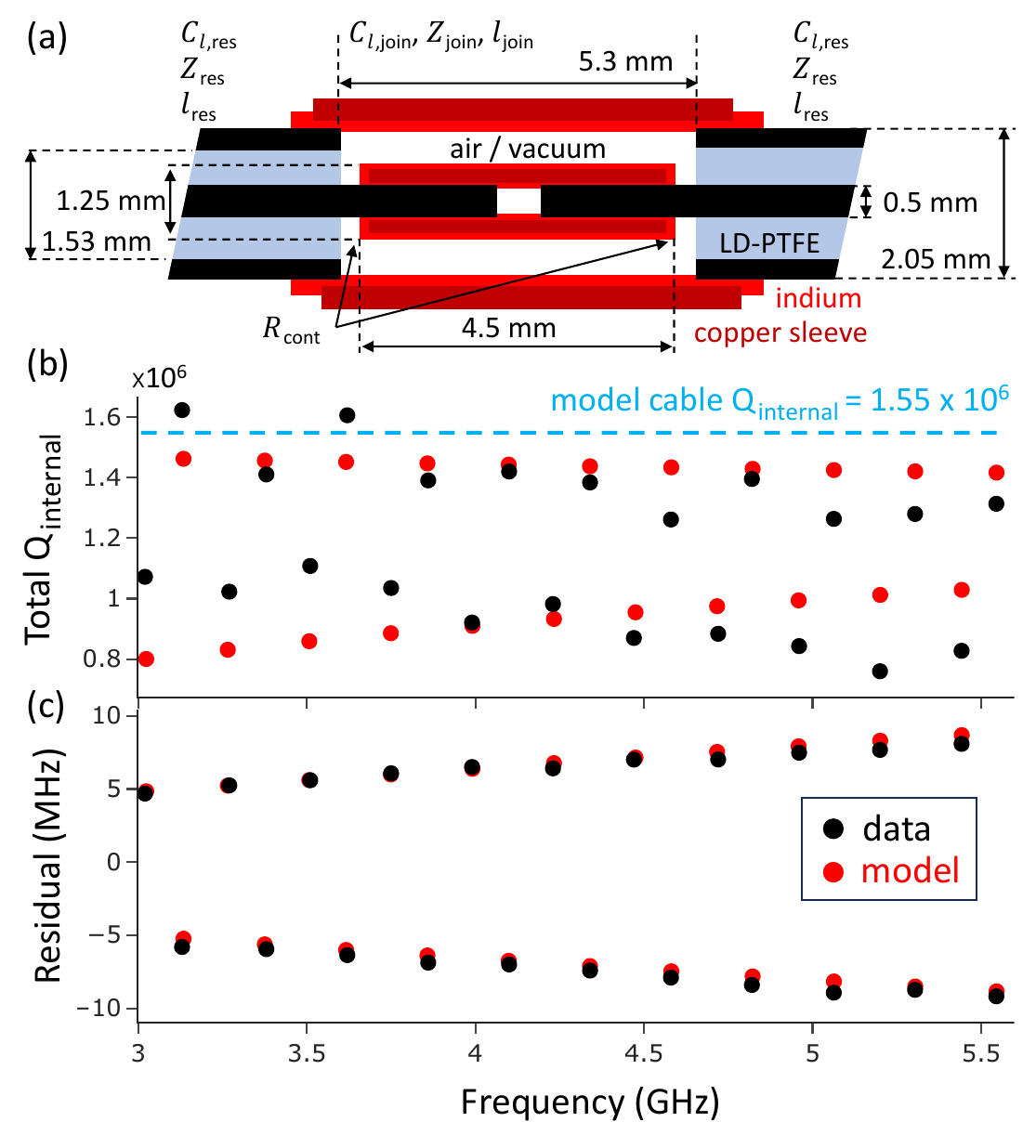}
\caption{(a) Diagram of the inner and outer conductor cable splice. (b) Experimental and ABCD matrix model total internal quality factors for the spliced cable resonator under test over the frequency range of 3 to 5.5 GHz. Resulting total mean internal quality factor for "odd" modes where the splice is located near a current node is $Q_{i} = 1.44 x 10^6$ for the model and $Q_{i} = 1.40 x 10^6$ for the experimental data. For the "even" modes where the splice is located near a current anti-node is $Q_{i} = 9.26 x 10^5$ for the model and $Q_{i} = 9.39 x 10^5$ for the experimental data. (c) Residual between observed resonance frequencies and a constant free spectral range of $120.75 MHz$ for model and experimental data.}
\label{fig:Splice_diagram}
\end{figure}

We extend this low-loss indium mechanical-intermixing method to both the inner and outer conductor of the coaxial line to form a splice in Fig. ~\ref{fig:Splice_diagram}. Here we use two copper tubes, coated in a thin indium layer, to form a bridge between the two aluminum cables. Preparation of the inner tube follows the same basic procedure as the outer tube described previously: indium is first soldered with flux and heat, and the excess is removed. As the inner indium-coated tube must fit tightly over the small diameter inner conductor of the cable and still leave a sufficient gap to the indium surface of the outer tube, the tube is necessarily very small diameter and thin. However as indium is very malleable, the excess indium on both the inner and outer surfaces of the tube can be readily trimmed, for example using a scalpel on the outer surface and a small drill bit or by pressing the coated tube into molds of the desired shape. Again, all these individual components can be prepared well in advance and rely purely on simple mechanical friction at room-temperature to create the low-loss join. Prior work has demonstrated a low loss splice by wirebonding machined cable end faces \cite{qiu2023deterministic}. In this work, the authors were able to achieve a reported cable mode quality factor of $1.39 x 10^6$ over a 64 meter link with 3 such joins. We study a splice by inserting the inner and outer conductor join shown in Fig. ~\ref{fig:Splice_diagram}(a) at the midpoint of a 1m cable resonator that is tested in the configuration of ~\ref{fig:reflection_measurement_diagram}(a). In this case, the splice falls near the current and voltage nodes for even and odd longitudinal mode orders respectively to study the effect of current and field dependent-losses.

To model this system, we can extend the ABCD matrix model of Equation \ref{eqn:cable} by replacing the cable $\bm{\mathit{T_{res}}}$ with two half cables, $\bm{\mathit{T_{A}}}$ and $\bm{\mathit{T_{B}}}$, a joint transmission line region $\bm{\mathit{T_{join}}}$, and a contact resistance for each interface $R_{cont}$:

\begin{equation}
    \bm{\mathit{T_{res}}}
    =
    \bm{\mathit{T_{A}}}
    \begin{bmatrix} 1 & R_{cont} \\ 0 & 1 \end{bmatrix}
    \bm{\mathit{T_{join}}}
    \begin{bmatrix} 1 & R_{cont} \\ 0 & 1 \end{bmatrix}
    \bm{\mathit{T_{B}}}
\end{equation}

Beyond fitting each individual resonance to fit the internal quality factor of each mode, we can then compare the extracted experimental quality factors against a single circuit model to compare quality factors across the studied frequency range as shown in Fig. ~\ref{fig:Splice_diagram}(b). Using the mean internal quality factor of the bare cable to set the cable loss, a fixed contact resistance of approximately $6x10^{-4} \Omega$ reproduces the trends in the even and odd mode data from the impact of the resistance relative to current nodes and anti-nodes. It is observed that the deviations from the fixed resistance model are frequency dependent, suggesting a higher effective contact resistance at higher frequencies. Comparing to prior literature on factory-fabricated mechanical splices, this push-to-connect indium splice is approximately one order of magnitude lower resistance \cite{renger2023cryogenic}. Finally, we note that the fit ABCD circuit model of the spliced cable resonator well predicts other experimental parameters such as the shifted frequencies of the even and odd modes due to the splice impedance discontinuities.

\begin{figure}
    \hspace*{-0.4cm}\includegraphics[width=0.54\textwidth]{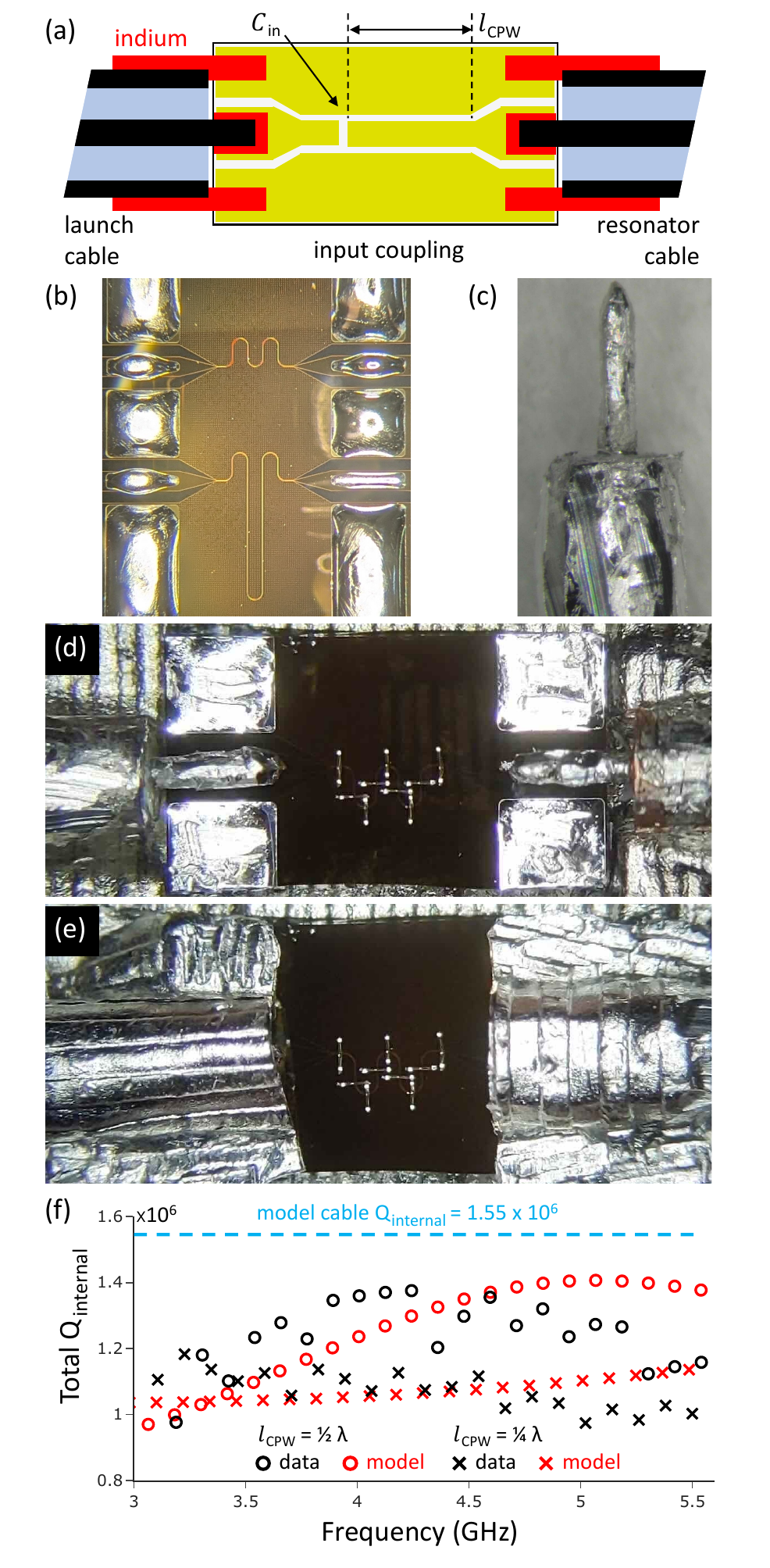}
	\caption{(a) Diagram of chip attach. (b) Micrograph of chip with indium solder reflowed on contact pads. (c) Indium-coated coaxial cable prior to assembly. (d) Micrograph of cable center conductor cold molded onto contact pad. (e) Micrograph of indium foil cold molded for ground to outer conductor connection. (f) Experimental and ABCD matrix model total internal quality factors over the frequency range of 3 to 5.5 GHz for the half-wavelength and quarter-wavelength chip-cable composite resonators.}
 
    
	\label{fig:chip_attach}
\end{figure}

In a similar vein, we can apply this low-loss indium mechnical-intermixing method to the cable-to-chip attach problem. We prepared a test chip comprising a set of cable-to-chip launch pads, small capacitances to form the boundaries of the cable-resonators-under-test, two different CPW trace lengths ($l_{CPW}$), and a second set of cable-to-chip launch pads (see Fig. ~\ref{fig:chip_attach}(a,b)). To better study loss sources at the interface between cable and chip, the two $l_{CPW}$ lengths correspond to a voltage-maximum and a current-maximum at the interface at 5GHz, half-wave and quarter-wave respectively. The launch pads of the test chip are patterned with a gold surface in the indicated regions on top of a standard superconductor-on-dielectric planar superconducting circuit \cite{rosenberg20173d}. In order to provide a suitable surface for indium cold-molding, at least $100 \mu m$ of indium must be added to these regions. This can be accomplished by manually reflowing indium solder balls in the contact pad region at approximately 180 \textdegree C for 10 s as shown in Fig. ~\ref{fig:chip_attach}(b). In this example assembly, 8 mg of indium was used on the central ground pads and 2 mg of indium was used on the centerline pads to create solder heights of $200 \mu m$ and $150 \mu m$ respectively.

Following standard methods, the center conductors of both launch and aluminum cables are exposed to the desired length ($\sim 2$mm) with the end-tip sharpened gently.  The same ultrasonic soldering method described previously is used on both the inner and outer conductors of the coaxial lines as shown in Fig. ~\ref{fig:chip_attach}(c). Similarly, these cable preparations can be done well in advance of cable usage, making this a manufacturable process. To achieve the final connection, we press the indium-coated center conductor into the indium on chip with tweezers and press-mold a thin indium foil between the indium on chip-ground and indium on coax outer conductor. To complete the assembly, mechanical strain relief is screwed into the chip-mount to hold the cables in place. 

The on-chip capacitor, co-planar waveguide and cable attach assembly replaces the air gap capacitor input coupling from Fig. ~\ref{fig:reflection_measurement_diagram}(a) to study the composite resonator performance of the bare cable, cable-to-chip attach, and short on-chip co-planar waveguide. Data from two test chips that were prepared with $100 \mu m$ indium on the contact pads is shown in Fig. ~\ref{fig:chip_attach}(b). The measured mean internal quality factors of the composite chip-cable resonators of the two test chips is $Q_i = 1.24 x 10^6$ for the case where the cable-to-chip attach is at a voltage-maximum and $Q_i = 1.07 x 10^6$ for the case where the cable-to-chip attach is at a current-maximum.

To model this system for current and voltage dependent loss mechanisms, we can extend the ABCD matrix model of Equation \ref{eqn:cable} by replacing the cable $\bm{\mathit{T_{res}}}$ with an on chip waveguide section $\bm{\mathit{T_{cpw}}}$, the cable region $\bm{\mathit{T_{cable}}}$, a contact resistance for the cable to chip attach $R_{cont}$, and a shunt resistance to account for any parasitic ohmic path or lossy dielectric absorption $R_{shunt}$:

\begin{equation}
    \bm{\mathit{T_{res}}}
    =
    \bm{\mathit{T_{cpw}}}
    \begin{bmatrix} 1 + \frac{R_{cont}}{2 R_{shunt}} & R_{cont} + \frac{R_{cont}^2}{4R_{shunt}} \\ \frac{1}{R_{shunt}} & 1 + \frac{R_{cont}}{2 R_{shunt}} \end{bmatrix}
    \bm{\mathit{T_{cable}}}
\end{equation}

By assuming an internal quality factor of the cable to be equal to the mean of the bare cable measurement as was done for the splice study, the contact and shunt resistances can be tuned to achieve mean model internal quality factors to match the two experimental data sets. The resulting $R_{cont} = 8.5 x 10^{-4}$ is higher, but comparable in magnitude to the cylinder splice example studied in Fig. ~\ref{fig:Splice_diagram}. Comparing to prior literature on wirebond cable-to-chip attach which achieved contact resistances in the range of $0.05 - 0.4 \Omega$ \cite{zhong2021deterministic,niu2023low}, this press-mold connection is approximately two orders of magnitude lower contact resistance. In contrast to the cylinder splice case however, a finite shunt resistance of $R_{shunt} = 1.3 x 10^{7}$ is necessary to explain the data beyond the internal quality factor of the bare cable, suggesting a field dependent loss mechanism present at the chip attach.

\begin{figure}
    \includegraphics[width=0.48\textwidth]{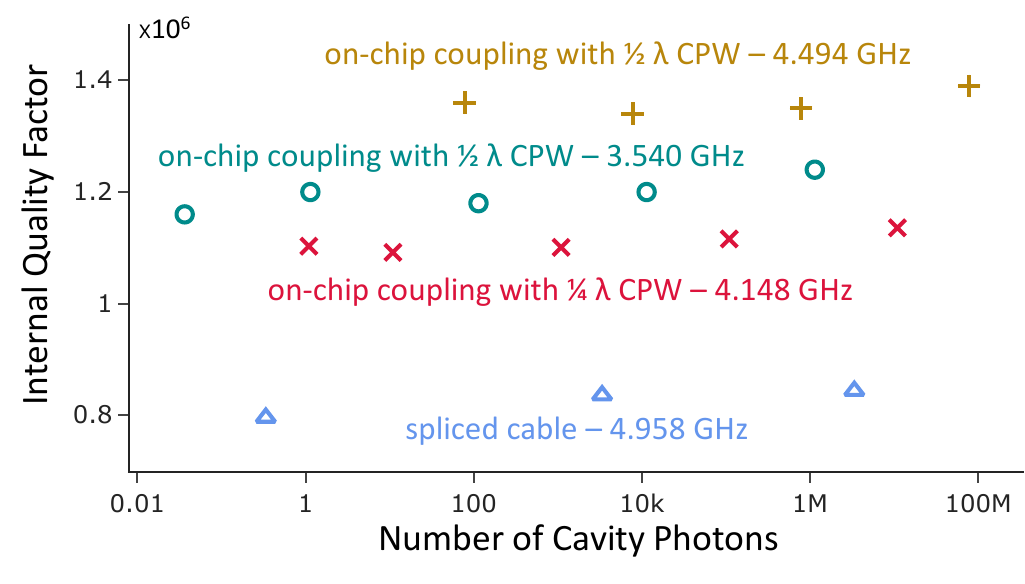}
	\caption{Power dependence of select experimental data from Figs. ~\ref{fig:chip_attach} and ~\ref{fig:Splice_diagram}.}
	\label{fig:power_dep}
\end{figure}

Since two level system (TLS) loss sources saturate with applied microwave power and the relevant losses for quantum applications are those that would affect single or few photon excitations, we must characterize the power dependence of the internal quality factors in our systems \cite{mcrae2020materials}. Conveniently for ease of measurement, we find only a very weak power dependence of extracted quality factors over the range of 1 to $10^8$ photons in the resonator for the variety of experimental conditions shown in Fig. ~\ref{fig:power_dep}. To calculate the photon number of the resonator under test for this dataset, the fractional absorbed power, $A$, must be calculated from the ratio of the internal and total quality factors:

\begin{equation}
    A = 1- \left( \frac{2 Q_t}{Q_i}-1 \right)^2
\end{equation}

The on resonance cavity photon number, $N_{photon}$ for a given input power at the circulator, $P_{in}$, is then determined by the input photon flux (calculated the Planck constant, $h$) multiplied by the intracavity photon lifetime:

\begin{equation}
    N_{photon} = \frac{Q_t P_{in} A}{2 \pi h f^2}
\end{equation}

For the experimental data presented in this work, the power dependence does not allow for the extraction of a meaningful trend of $Q_i$ vs. photon number. The variation in internal quality factor is commensurate with error in fitting the data. Further studies of more resonances over a larger range of powers are needed to understand the physical origin of any residual power dependence. Comparing to prior literature, this observation of no major trend with measurement power is similar to the prior bare cable characterization report \cite{kurpiers2017characterizing} and in contrast to prior studies that included on-chip transmission lines for characterization \cite{zhong2021deterministic,niu2023low,qiu2023deterministic}.

To conclude, we have presented a variety of techniques based on a mechanically-intermixed indium method for the characterization, splice, and chip-attach of superconducting coaxial cables. The presented data is based on a single cable-type from one vendor. Further study is needed to understand if these results represent a fundamental limit of available materials or if further improvements in the internal quality factor of superconducting cables in the single-photon, mK temperature regime are possible to enable even higher performance quantum interconnects for future superconducting quantum systems. We also envision that the push-to-connect nature of these joins can be leveraged for on-chip connectors that enable simple, modular assembly of future superconducting quantum systems.

\end{document}